\documentclass[%
superscriptaddress,
showpacs,preprintnumbers,
nofootinbib,
amsmath,amssymb,
aps,
twocolumn,
]{revtex4}
\usepackage{braket}
\usepackage{bm}
\usepackage{dcolumn}
\usepackage{hyperref}
\usepackage[final]{graphicx}
\usepackage{ulem}
\input{colordvi.tex}
\usepackage{color}
\newcommand{\G}{\,{\rm G}}
\newcommand{\Mpc}{\,{\rm Mpc}}

\newcommand{\eV}{\,{\rm eV}}
\newcommand{\keV}{\,{\rm keV}}
\newcommand{\GeV}{\,{\rm GeV}}
\newcommand{\bea}{\begin{eqnarray}}
\newcommand{\eea}{\end{eqnarray}}
\begin{document}
\title{Late-time magnetogenesis driven by ALP dark matter and dark photon}

\author{Kiwoon Choi}
\email{kchoi@ibs.re.kr}
\affiliation{Institute for Basic Science, Center for Theoretical Physics of the Universe, Daejeon 34051, South Korea}

\author{Hyungjin Kim}
\email{hyungjin.kim@weizmann.ac.il}
\affiliation{Department of Particle Physics and Astrophysics, Weizmann Institute of Science, Rehovot 7610001, Israel}
\affiliation{Institute for Basic Science, Center for Theoretical Physics of the Universe, Daejeon 34051, South Korea}

\author{Toyokazu Sekiguchi}
\email{sekiguti@resceu.s.u-tokyo.ac.jp}
\affiliation{Research Center for the Early Universe (RESCEU), Graduate School of Science, The University of Tokyo, Tokyo 113-0033, Japan}
\affiliation{Institute for Basic Science, Center for Theoretical Physics of the Universe, Daejeon 34051, South Korea}
\date{\today}

\begin{abstract}
We propose a mechanism generating primordial magnetic fields after the $e^+e^-$ annihilations. Our mechanism involves an ultra-light axion-like particle (ALP) which constitutes the dark matter, and a dark $U(1)_X$ gauge boson introduced to bypass the obstacle placed by the conductivity of cosmic plasma. In our scheme, a coherently oscillating ALP amplifies the dark photon field, and part of the amplified dark photon field is concurrently converted to the ordinary magnetic field through the ALP-induced magnetic mixing. For the relevant ALP mass range $10^{-21} {\rm eV}\lesssim m_\phi\lesssim 10^{-17}{\rm eV}$, our mechanism can generate $B\sim 10^{-24} \,{\rm G} \,(m_\phi/10^{-17} {\rm eV})^{5/4}$ with a coherent length $\lambda \sim (m_\phi/10^{-17} {\rm eV})^{-1/2}$ kpc, which is large enough to provide a seed of the galactic magnetic fields. The mechanism also predicts a dark $U(1)_X$ 
electromagnetic field $E_X \sim B_X\sim 80\,{\rm nG}\, (m_\phi/10^{-17}{\rm eV})^{-1/4}$, which can result in interesting astrophysical/cosmological phenomena by inducing the mixings between the ALP, ordinary photon, and dark photon states. 
\end{abstract}

\preprint{CTPU-PTC-18-04, RESCEU-4/18}


\maketitle

The origin of the primordial magnetic fields is one of the 
longstanding problems in modern cosmology. 
In a variety of cosmological scales, magnetic fields are observed\,\cite{Kronberg:1993vk,Grasso:2000wj,Durrer:2013pga}.
For instance, radio observations have revealed that
magnetic fields of $\mathcal O(1\textrm{--}10) \mu$G are inherent
in the galaxies and clusters of galaxies, which might  originate from a primordial seed field\,\cite{Davis:1999bt} 
\bea
B_{\rm seed}\gtrsim  {\cal O}(10^{-30})\,  \G\,\,\,\, {\rm with}\,\,\,\, \lambda \gtrsim 0.1 \, {\rm kpc},
\eea 
which is amplified later by the dynamo mechanism~\cite{Brandenburg:2004jv}, where $\lambda$ denotes the coherent length of the corresponding $B$ fields.
More recently, the existence of magnetic fields in cosmic voids has been
inferred from the observations of TeV blazars\,\cite{Neronov:1900zz,Tavecchio:2010mk,Dolag:2010ni,
Essey:2010nd,Taylor:2011bn,Takahashi:2013lba,Finke:2015ona}.
Those observations have found a lack of secondary gamma-rays in the GeV range, which
ought to be emitted by the electrons/positrons produced from the collision of the 
primary gamma-rays with extragalactic background light. This can be explained if there exist magnetic fields at intergalactic voids
\begin{equation}
B_{\rm void}\times{\rm min}\left[1,\sqrt{\lambda/0.1\Mpc}\right]\gtrsim {\cal O}(10^{-19}\textrm{--}10^{-16})\G,
\end{equation}
which deflect the trajectory of the produced electrons/positrons away from the direction of the primary gamma-rays \,\cite{Durrer:2013pga}.

So far, a number of models are proposed for  cosmological magnetogenesis in the early Universe
(for a recent review, see e.g. \cite{Durrer:2013pga}). 
An interesting and extensively studied  possibility is the inflationary magnetogenesis scenario \cite{Turner:1987bw,Ratra:1991bn,Demozzi:2009fu,Barnaby:2012tk,Fujita:2012rb,Fujita:2016qab,Adshead:2016iae}. In this scenario, magnetic fields created inside the horizon can be 
stretched out to superhorizon scales and can have a comoving correlation length 
comparable to the current cosmological scales.
However, inflationary magnetogenesis often
suffers from the backreaction by the produced magnetic fields, which spoils the dynamics of inflaton or  generates too large non-Gaussianity in primordial perturbations\,\cite{Demozzi:2009fu,Barnaby:2012tk,Fujita:2012rb}.\footnote{These problems can be circumvented if magnetogenesis takes place after the CMB scales exit the horizon\,\cite{Fujita:2016qab,Adshead:2016iae}.}
There are other magnetogenesis scenarios, for instance, based on  phase transition in the early Universe\,\cite{Vachaspati:1991nm,Enqvist:1993np}; however, those scenarios are still lacking  concrete realization.\footnote{See also~\cite{Miniati:2017kah} for a generation of primordial magnetic field with QCD axion prior to the QCD phase transition.}

Another interesting but less explored possibility is a late-time magnetogenesis which takes place well after big bang nucleosynthesis (BBN). 
Such a late-time scenario would have better prospects to give a large coherence length of the produced $B$ fields and also may come up with concrete predictions, as the early Universe after the BBN is highly constrained. 
In this paper, we propose a novel mechanism of late-time magnetogenesis, which occurs after the $e^+e^-$  annihilations.

Our mechanism involves an ultralight axion-like particle (ALP) $\phi$ and a dark $U(1)_X$ gauge field $X_\mu$, whose dynamics is described by the Lagrangian 
\begin{eqnarray}
\label{lagrangian_eff}
\mathcal L&=& \frac12\partial_\mu\phi\partial^\mu\phi - \frac12m_\phi^2 \phi^2
-\frac14F_{\mu\nu}F^{\mu\nu}-\frac14X_{\mu\nu}X^{\mu\nu} \notag \\
&&
-\frac{g_{AA}}{4f}\phi F_{\mu\nu}\widetilde F^{\mu\nu}
-\frac{g_{XX}}{4f}\phi X_{\mu\nu}\widetilde X^{\mu\nu}
-\frac{g_{AX}}{2f}\phi F_{\mu\nu}\widetilde X^{\mu\nu} \notag \\
&&+J^\mu A_\mu,
\label{eq:lagr}
\end{eqnarray}
where $F_{\mu\nu}=\partial_\mu A_\nu-\partial_\nu A_\mu$ and $X_{\mu\nu}=\partial_\mu X_\nu-\partial_\nu X_\mu$ 
are the $U(1)_{\rm em}$ and $U(1)_X$ field strengths, $\widetilde F_{\mu\nu}$ and $\widetilde X_{\mu\nu}$ are their duals,
$J^\mu$ is the standard electromagnetic current,  and $f$ is a dimensionful parameter describing  the initial ALP misalignment: \bea
f\equiv \phi_{\rm initial}.
\eea
In the expanding Universe with the metric
\bea
ds^2 = a^2(\tau)(d\tau^2-d{\bm x}^2),
\eea the equations of motion for the ALP and $U(1)$ gauge bosons are given by
\bea
\label{eqn_motion}
&&\ddot\phi+2{\cal H}\dot\phi-\nabla^2\phi +a^2m_\phi^2\phi   
= -\frac{1}{a^2}\left(\frac{g_{AA}}{f}\dot{\bm A}\cdot \nabla\times {\bm A} \right.\nonumber \\
&&  \left.+\frac{g_{XX}}{f}\dot{\bm X}\cdot \nabla\times {\bm X} 
+\frac{g_{AX}}{f}(\dot{\bm A}\cdot \nabla\times {\bm X} +\dot{\bm X}\cdot \nabla\times {\bm A} )  \right), \nonumber \\
&&\ddot{{\bm A}} + \sigma\left(\dot{\bm A}+{\bm v}\times (\nabla\times {\bm A})\right)+ \nabla\times (\nabla\times {\bm A})\nonumber \\&&\hskip -0.5cm = \frac{g_{AA}}{f}\left(\dot\phi \nabla\times{\bm A}-\nabla\phi\times \dot{\bm A}\right)+\frac{g_{AX}}{f}\left(\dot\phi \nabla\times{\bm X}-\nabla\phi\times \dot{\bm X}\right),
\nonumber \\
&&\ddot{{\bm X}} + \nabla\times (\nabla\times {\bm X})=\frac{g_{XX}}{f}\left(\dot\phi \nabla\times{\bm X}-\nabla\phi\times \dot{\bm X}\right)\nonumber \\
&&\hskip 2cm +\frac{g_{AX}}{f}\left(\dot\phi \nabla\times{\bm A}-\nabla\phi\times \dot{\bm A}\right), \eea
where we used the temporal gauge $A_\mu=(0,\bm A), X_\mu=(0, \bm X)$. 
Here, the dots denote the derivatives with respect to the conformal time $\tau$, 
${\cal H}=\dot a/a=aH$ is the conformal Hubble expansion rate, and finally, Ohm's law
${\bm J}=\sigma ({\bm E}+{\bm v}\times {\bm B})$ is used for the equation of ${\bm A}$, where $\sigma$ is the conformal conductivity of the cosmic plasma, and ${\bm v}$ is a fluid velocity field.

The following is a brief summary of how our mechanism works and what the underlying assumptions are. 
At $\tau_{\rm osc}$ when $3H(\tau_{\rm osc})\simeq m_\phi$, the ALP $\phi$ commences to oscillate to form the dark matter.
Coherently oscillating $\phi$  causes a tachyonic instability of $X_\mu$ through the coupling $g_{XX}$ and amplifies the dark photon field strength as
\bea
\label{B_X-field}
E_X \simeq B_X \simeq 83 \,{\rm nG}\, (m_\phi/ 10^{-17}\eV )^{-1/4}
\eea
for $g_{XX} \gtrsim {\cal O}(10)$.
For an efficient amplification, we assume that $X_\mu$  is strictly massless and there is no light $U(1)_X$-charged particle. 
In the presence of the magnetic mixing coupling $g_{AX}$~\cite{Kaneta:2016wvf}, the amplified $E_X\sim B_X$ are partly converted into the ordinary magnetic fields, generating
\bea
\label{B-field}
B\simeq 1.7 \times 10^{-24} \,{\rm G}\,  (m_\phi/ 10^{-17}\eV )
\left(\frac{g_{AX}/f}{10^{16}\GeV}\right),
\eea
with a coherent length $\lambda \sim  (m_\phi/10^{-17}{\rm eV})^{-1/2}$ kpc. 
The conversion of $X_\mu$ to $A_\mu$ is  most efficient when the conductivity $\sigma \propto n_e$  is minimized, which happens when the electrons/positions are annihilated enough, so the electron density is suppressed as $n_e \sim n_{\rm baryon}\sim 10^{-9}n_\gamma$. 
This happens when $T\lesssim 20$ keV.
For this reason, we consider the ALP mass range
\bea 
10^{-21} {\rm eV}\lesssim m_\phi \lesssim 10^{-17} {\rm eV}
\label{mphi_range}
\eea
 for which ALP begins to oscillate at $T_{\rm osc} \simeq 100 \, \keV \times (m_\phi /10^{-17} \eV)^{1/2}$, so that magnetogenesis occurs at $T\sim T_{\rm osc}/5\lesssim 20 \, {\rm keV}$ as desired. 
Here  the lower bound on $m_\phi$ is imposed to be compatible with the Lyman-$\alpha$ constraint on ultralight ALP dark matter~\cite{Irsic:2017yje}.
Then the temperature range of our mechanism is 
\bea
\label{T_range}
200 \, {\rm eV}\lesssim T \lesssim 20 \, {\rm keV},\eea
for which the conductivity is determined by the Thomson scattering as\,\cite{Takahashi:2007ds,Giovannini:2009bk,Hollenstein:2012mb} 
\bea
\label{conduc}
\sigma_{\rm phy} =\frac{\sigma}{a} \simeq \frac{135\zeta(3)}{e^2 \pi^3}\frac{m_e^2}{T} \frac{n_e}{n_\gamma}. 
\eea

Our model (\ref{lagrangian_eff}) generically involves   
three ALP couplings:  $g_{AA},~g_{AX}$, and $g_{XX}$. 
To check the feasibility of our mechanism, we list current observational bounds on the ALP couplings.\footnote{Our model obviously satisfies the recent experimental bounds on ALP couplings from nuclear spin process \cite{Abel:2017rtm} as in our case
the quantized ALP-gluon coupling is exactly zero and the derivative couplings of ALP to nucleons, which are radiatively induced by $g_{AA}$, are much smaller than the bound of  \cite{Abel:2017rtm}.}
For an ultralight ALP, major constraints on $g_{AA}$ come from  astrophysical observations based on the photon-ALP conversion  \cite{Raffelt:1987im}, e.g. X-ray observations\,\cite{Wouters:2013hua,Berg:2016ese,Conlon:2017qcw,Chen:2017mjf,Marsh:2017yvc}, quasar spectra\,\cite{Ostman:2004eh}, cosmological tests of the distance-duality relation\,\cite{Avgoustidis:2010ju,Tiwari:2016cps}, and CMB spectral distortions\,\cite{Mirizzi:2005ng,Tashiro:2013yea,Mirizzi:2009nq,Mukherjee:2018oeb}. In our case,
the strongest bound on $g_{AA}$ comes from X-ray observation~\cite{Conlon:2017qcw} yielding $g_{AA} /f  \lesssim 1.5\times 10^{-12} \GeV^{-1}$.
Constraints on $g_{AX}$ can be drawn  also from the ALP-photon mixing induced by background  $B_X$. We find
the strongest bound on $g_{AX}$ in our case comes from the CMB spectral distortion yielding~\cite{Tashiro:2013yea}%
\footnote{Note that the bound of ~\cite{Tiwari:2012ig} based on WMAP observation does not apply for our case
as ~\cite{Tiwari:2012ig} assumes that ${\bm B}$ fields have a power law spectrum and
coherent length $\sim$ Mpc, which are not shared by the dark photon field strength in our scenario. 
}
\bea
\label{bound_magnetic_mixing}
\frac{g_{AX}}{f} \lesssim  10^{-16} \GeV^{-1} \left( \frac{B_X}{100 \, {\rm nG}} \right)^{-1}.
\eea
The coupling $g_{XX}/f$  is far less constrained as it involves only the dark photon fields, and can have a value large enough to implement our mechanism.
   
To motivate the introduction of  $X_\mu$, let us briefly discuss the generation of magnetic field in the absence of $X_\mu$. 
When $\sigma_{\rm phy}\gg m_\phi$, we can approximate the equation of motion for the photon field as $\sigma \dot{A}_k \approx g_{AA} k \dot{\phi} A_k / f$ in the Fourier space, where $k$ is the comoving wave number.
This allows a solution for the magnetic field as $A_k = A_{k,\rm vac} \exp [ \int d\tau' \, \Gamma]$ where $A_{k,\rm vac}$ is the vacuum fluctuation of the gauge field, and $\Gamma =g_{AA} k \dot{\phi}  / (\sigma f)$ is the magnetic field production rate. 
Then the magnetic field production for the unit Hubble time at $\tau\sim \tau_{\rm osc}$ is estimated as
\bea
\frac{1}{\cal H}\frac{\dot {\bm A}}{\bm A} 
= \frac{\Gamma}{\cal H}
\lesssim \left(\frac{g_{AA}}{f}\right)^2\frac{m_\phi f^2}{\sigma_{\rm phy}}\lesssim  5\times 10^{-9}
\eea
where $k$ is within the instability regime, i.e. $k/a\lesssim {\cal O}(g_{AA}m_\phi)$. Here, we used $\dot \phi/{\cal H} f={\cal O}(1)$ at $\tau\sim \tau_{\rm osc}$ and $\sigma_{\rm phy}$ given by (\ref{conduc}), together with the bounds on $g_{AA}/f$ and $m_\phi$,  while assuming  
$f\lesssim M_{\rm Pl}$ which is necessary to avoid a too large relic mass density of $\phi$.  
The above production rate is too weak to yield any appreciable amount of $B$ fields, which is essentially due to the huge suppression by $m_\phi/\sigma_{\rm phy}\lesssim 2\times 10^{-17}$.

With the dark photon field $X_\mu$, we can make  magnetogenesis much more efficient.
As the cosmic plasma is neutral to $U(1)_X$, $X_\mu$ can be freely  amplified by the tachyonic instability caused by the oscillating $\phi$, and this exponential amplification can compensate for much of the suppression by $m_\phi/\sigma_{\rm phys}$.   In the following, we describe our magnetogenesis mechanism  and present some of the key results, while leaving more detailed study to the forthcoming work \cite{detail}.

Right after $\tau_{\rm osc}$, when the energy density $\rho_X$ of $X_\mu$ is negligible compared to $\rho_\phi$,  $\phi$
evolves as 
\begin{equation}
\theta\equiv \frac{\phi}{f}\approx \left(\frac{a(\tau)}{a(\tau_{\rm osc})}\right)^{-3/2} \cos[m_\phi (t-t_{\rm osc})],
\label{eq:phi0}
\end{equation}
where $t=a(\tau)\tau/2$.
In this stage, $\phi$ is approximately homogeneous, and the backreaction from $A_\mu$ can be ignored. Then the equation of motion of $X_\mu$ in the momentum space is approximated by 
\bea
\label{instability_X}
\ddot {\bm X}_{k\pm} + k(k\mp g_{XX}\dot\theta) {\bm X}_{k\pm}\simeq 0,
\eea
where the subscript $\pm$ denotes the helicity.
This shows that under the oscillating $\phi$, one of the helicity states of ${\bm X}_k$ experiences a tachyonic instability for  certain range of $k$, and the vacuum fluctuations of ${\bm X}_k$ in this range of $k$ are exponentially amplified to be a stochastic classical field.
   
At a certain time $\tau_X>\tau_{\rm osc}$, $\rho_X$ becomes comparable to $\rho_\phi$, where $\tau_X/\tau_{\rm osc}=a(\tau_X)/a(\tau_{\rm osc})$ is determined mostly by the coupling $g_{XX}$. 
Around this time, the initial energy density of the zero momentum mode of $\phi$ is converted mostly to $\rho_X$, and also partly to the energy density of nonzero momentum modes of $\phi$. 
As was shown in \cite{Kitajima:2017peg}, the dark photon field production is particularly efficient for $g_{XX}\gtrsim{\cal O}(10)$, and in this paper we will use $g_{XX}=100$ as a benchmark point for explicit analysis.
The ordinary electromagnetic field $A_\mu$ is produced also around this time by the magnetic mixing coupling $g_{AX}$. 
As the conductivity dominates over other factors, the production is described by the following approximate equation of motion
\bea
\label{eqn_A}
\sigma \dot {\bm A}\simeq g_{AX}\left(\dot \theta \nabla\times {\bm X} -\nabla\theta \times \dot {\bm X}\right),
\eea
where the effects of non-zero momentum modes of $\phi$ are included.

For a more quantitative analysis, we define
\bea
\label{ratios}
r(\tau)&\equiv & \frac{\langle \rho_\phi\rangle(\tau)}{(\rho_\phi)_{g=0}(\tau)}, \nonumber \\
 \epsilon (\tau) &\equiv & \frac{a(\tau)}{a(\tau_{\rm osc})}\frac{\langle \rho_X\rangle(\tau)}{(\rho_\phi)_{g=0}(\tau)}, \nonumber \\
 b(\tau) &\equiv &  
 \frac{1}{g_{AX}}
 \frac{\sqrt{\langle \rho_A\rangle(\tau)}}{ \sqrt{\langle\rho_X\rangle(\tau)}},
\eea
where $\langle\rho\rangle$ denotes the spatially averaged energy density, and $(\rho_\phi)_{g=0}$ is the homogeneous energy density of $\phi$ in the absence of gauge field production, i.e.
when $g_{AA}=g_{XX}=g_{AX}=0$. 
The backreaction from $A_\mu$ can be safely ignored
for $g_{AA}$ and $g_{AX}$ satisfying the observational bounds. 
Then  the evolutions of $r$ and $\epsilon$ are 
determined mostly by $g_{XX}$, while being insensitive to other model parameters. On the other hand, as we will see below, the evolution of $b(\tau)$ depends significantly on $m_\phi$.  Obviously, in the early stage at $\tau <  \tau_X$,  $r\simeq 1$, and $\epsilon$ is negligibly small.  
In the intermediate stage at $\tau\simeq \tau_X$, $r$ drops to a value which is an order of magnitude smaller than unity, while $\epsilon$ rises to a value of order unity. 
In the final stage at $\tau\gg \tau_X$, the three fields $\phi, X_\mu$, and $A_\mu$ are decoupled from each other and freely evolve. 
As a result, the energy densities evolve as $\langle\rho_\phi\rangle \propto 1/a^3$ and $\langle \rho_{X,A}\rangle\propto 1/a^4$, and $r(\tau), \epsilon(\tau)$, and $b(\tau)$ all approach some constants.

As $X_\mu$ is exponentially amplified by the coupling $g_{XX}$, it strongly back-reacts to the evolution of $\phi$, and develops an inhomogeneous part of $\phi$ for $\tau \gtrsim \tau_X$. 
A lattice simulation is required for a quantitative analysis of the evolution of our system.
Yet, the dependence of the final results  on  $m_\phi$ and $f$ can be determined by simple dimensional analysis. For this, let us first note that $r$, $\epsilon$, and also $a(\tau_{\rm osc})/a(\tau_X)$
are  insensitive\footnote{For $g_{XX}\gtrsim {\cal O}(10)$,
the initial $\rho_\phi$ is abruptly converted to $\rho_X$  at $\tau\sim \tau_X$, which is determined mostly by $g_{XX}$. After this abrupt conversion, $\langle\rho_\phi\rangle$ 
  and $a(\tau)\langle \rho_X\rangle$  evolve like $(\rho_\phi)_{g=0}\propto 1/a^3(\tau)$. As a consequence, their ratios, i.e. $r$ and $\epsilon$, are insensitive to $m_\phi$ and $f$, while there can be a logarithmic dependence which will be ignored here.} to  $m_\phi$ and $f$.
We then find the following simple power-law dependences of the relevant quantities on $m_\phi$ and $f$,
\bea
\label{parametric}
&&  a(\tau_X) \propto \tau_X\propto  1/T_X \propto a(\tau_{\rm osc}) \propto m_\phi^{-1/2}, \nonumber \\
&& B_X^2 \propto a^4\langle\rho_X\rangle(\tau_X)\propto  a^4 \langle\rho_\phi\rangle (\tau_X) \propto a^4(\tau_X)m_\phi^2 f^2 \propto f^2,\nonumber 
\\
&& k_* \sim g_{XX}\dot\theta(\tau_X) \propto a(\tau_X) m_\phi \propto m_\phi^{1/2} ,\nonumber \\
&&
\sigma(\tau_X) = a(\tau_X) \sigma_{\rm phy}(\tau_X)  \propto a(\tau_X)/T(\tau_X) \propto m_\phi^{-1}, 
\eea
where $B_X^2 = \langle |\nabla \times {\bm X}|^2 \rangle$,  and
 $k_*$ denotes the characteristic wave number of the produced  $X_\mu$ and $A_\mu$. 
One can also infer from (\ref{eqn_A}) that  
\bea
\label{b-parametric}
b(\tau_X) \propto 
\tau_X\dot \theta(\tau_X) k_*/\sigma(\tau_X) \propto m_\phi^{3/2},\eea
where we used the parametric dependences listed in (\ref{parametric}).

From (\ref{ratios}), the relic mass density of $\phi$ is determined as
\bea 
\label{alp_density}
\Omega_\phi h^2 = r(\tau_0) (\Omega_\phi h^2)_{g=0}
\simeq   0.5 \,\,
r(\tau_0) m_{-17}^{1/2} f_{16}^2, 
\eea
where $(\Omega_\phi h^2)_{g=0}$ is the relic density in the absence of gauge field production, $\tau_0$ is the conformal time at present, and  $m_{-17} \equiv m_\phi/ 10^{-17} {\rm eV}$, $f_{16}\equiv f/10^{16}{\rm GeV}$.
The produced dark photon field and its energy density can be parametrized as
\bea
 B_X(\tau_0) &\simeq& 
21\,{\rm nG} \, \times
 \sqrt{ \frac{\epsilon(\tau_0)}{r(\tau_0)} \frac{\Omega_\phi h^2}{m_{-17}^{1/2}}}
 \\
\Delta N_{\rm eff} &\simeq &
3.6\times 10^{-4} \,\,
 \frac{\epsilon(\tau_0)}{r(\tau_0)} \frac{\Omega_\phi h^2}{m_{-17}^{1/2}}.
\label{eq:omphi}
\eea
Taking the ALP mass dependence of $b$ in (\ref{b-parametric}), we can parametrize also the present value of the produced $B$ fields as 
\bea
B(\tau_0) =
(3\times 10^{-8} {\rm G}) \,
 \bar{g}_{-16}  m_{-17}  \Omega_\phi h^2
 \frac{\sqrt{\epsilon(\tau_0)}}{r(\tau_0)}
 \left(\frac{b(\tau_0)}{m_{-17}^{3/2}}\right),
 \eea
where $\bar{g}_{-16} \equiv g_{AX} / f_{16}$. 
Finally, from the instability equation  (\ref{instability_X}),  the characteristic size of the wave numbers of the dark matter $\phi$ and dark radiation $X_\mu$ can be estimated as 
 \bea
 \label{k*}
 k_*\sim g_{XX} \dot\theta \sim  g_{XX}\left(\frac{a(\tau_{\rm osc})}{a(\tau_X)}\right)^{1/2}a(\tau_{\rm osc}) m_\phi.
 \eea

\begin{figure}
\centering
\includegraphics[scale=0.55]{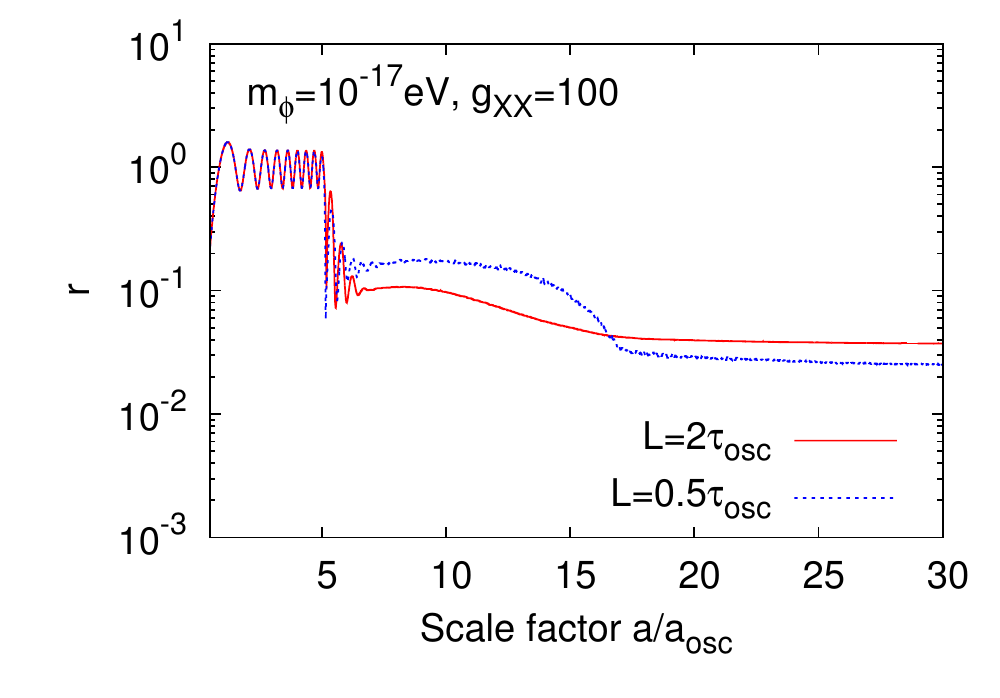} 
\includegraphics[scale=0.55]{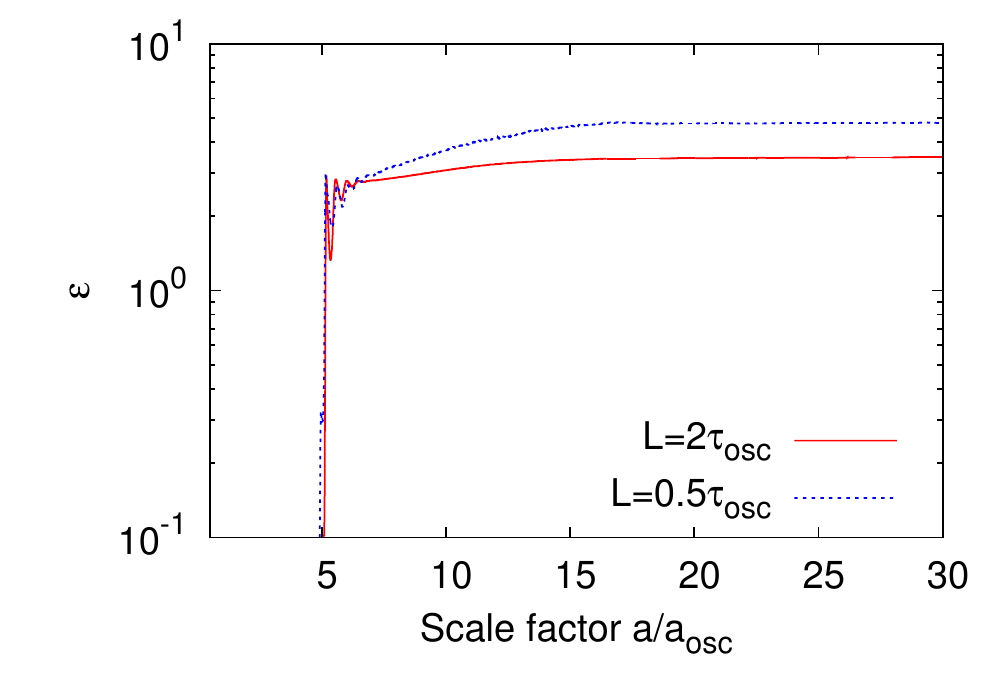} 
\includegraphics[scale=0.55]{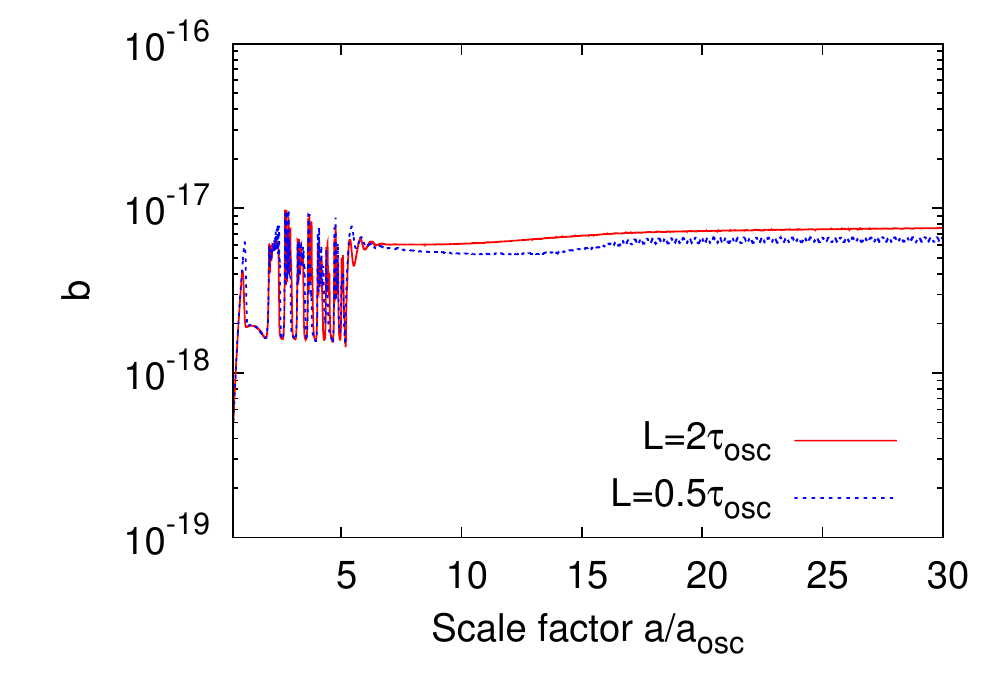} 
\caption{
 The time evolution of $r$ (top), $\epsilon$ (middle), and $b$ (bottom) from the lattice calculation with the number of grid $128^3$.
 One can also see $a_X/a_{\rm osc}\simeq 5$ from the epoch when $\epsilon$ and $b$ almost saturate. 
 Note that two different simulation boxes with comoving side lengths $L=2\tau_{\rm osc}$ (red solid) and $L=0.5\tau_{\rm osc}$ (blue dotted)
 show consistent results.}
 \label{fig:numerical}
 \end{figure}
 
 \begin{figure}
 \centering
 \includegraphics[scale=0.7]{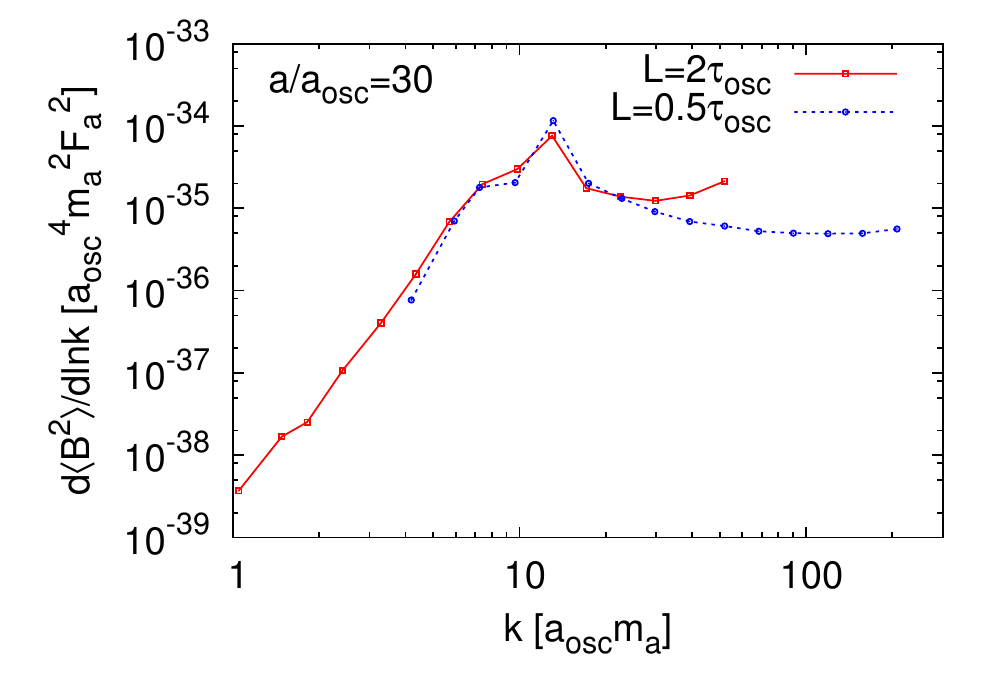}
 \caption{
 The power spectrum of produced magnetic fields from the lattice calculation at $a/a_{\rm osc}=30$. 
 Setup is the same as in Fig.~\ref{fig:numerical}. 
Results from different simulation boxes are consistent around the spectral peak.}
 \label{fig:spectrum}
 \end{figure}

Following \cite{Kitajima:2017peg}, we performed lattice calculations to examine the evolution of $r(\tau), \epsilon(\tau)$, and $b(\tau)$
for the benchmark point with $g_{XX}=100$. When performing the lattice simulation, we ignore the evolution due to magneto-hydrodynamics, i.e. ${\bm v} =0$, which we will discuss shortly. Taking various different values of $m_\phi$ and $f$, we confirmed that $r$ and $\epsilon$ are indeed insensitive to $m_\phi$ and $f$, while  $b$ depends on $m_\phi$ as (\ref{b-parametric}). In Fig.~\ref{fig:numerical}, we depict the results for $g_{XX}=100$ and $m_\phi=10^{-17}$ eV, showing $a(\tau_X)/a(\tau_{\rm osc})\simeq 5$ and the asymptotic values  $r\simeq 0.03,\, \epsilon\simeq 4, \, b\simeq 7 \times 10^{-18}$ at $\tau\gg \tau_X$. 
The power spectrum of the magnetic field $B$ is plotted in Fig.~\ref{fig:spectrum}, which exhibits a single spectral peak.%
\footnote{The spectrum with larger box size (solid red) exhibits fake enhancement at the cut-off wave number. 
This is caused by the mode-mode coupling of the dark photon and ALP fluctuations, which transfers energy towards high wave numbers.}
From those results, we find the dark and ordinary magnetic fields given in \eqref{B_X-field} and \eqref{B-field} with the correlation length 
\bea
\label{length}
\lambda &=& \frac{2\pi}{k_*} \,\simeq \, 0.3\, {\rm kpc}
 \,\, m_{-17}^{-1/2}.
\eea
The above correlation length is obtained from Fig.~\ref{fig:spectrum} and is about three times larger than the naive estimation  (\ref{k*}).
The produced dark and ordinary magnetic fields contribute to the effective number of relativistic degrees of freedom as
\bea
\Delta N_{\rm eff} &\simeq &  6\times 10^{-3}  m_{-17}^{-1/2},
\eea
which would be consistent (up to $2\sigma$) with the current observation $N_{\rm eff} =3.15 \pm 0.23 $~\cite{Ade:2015xua} for $m_\phi\gtrsim  10^{-21}$ eV.

Let us comment on the evolution of $B$ after the production.
We expect that the magnetic fields are frozen-in after the production.
The $B$ fields at kpc scales do not 
dissipate away even after the recombination due to the high conductivity of the Universe. On the other hand, the Alfv\'en crossing time, which sets the interaction time scale between magnetic fields and plasma\,\cite{Banerjee:2004df}, is much larger than the age of the Universe. Thus it is expected that the $B$ fields are not subject to an evolution due to magneto-hydrodynamics.

In our scenario, the dark photon gauge field strengths are inevitably produced on cosmological scales. 
The produced $B_X\sim E_X$ can induce a mixing between the ALP, ordinary photon, and dark photon states, 
which may result in interesting astrophysical/cosmological consequences \cite{Wouters:2013hua,Berg:2016ese,Conlon:2017qcw,Chen:2017mjf,Marsh:2017yvc,Ostman:2004eh,Avgoustidis:2010ju,Tiwari:2016cps,Mirizzi:2005ng,Tashiro:2013yea,Mirizzi:2009nq,Mukherjee:2018oeb,Ejlli:2016asd}.  Under $B_X$ given in (\ref{B_X-field}),  we find the bound (\ref{bound_magnetic_mixing}) is translated into $g_{AX}\lesssim 5$ when $\phi$ constitutes the dark matter in the Universe.  As was noticed in \cite{Tashiro:2013yea}, future measurements of CMB distortions by PIXIE/PRISM can improve the bound on $g_{AX}$ by two orders of magnitude. 
This implies that PIXIE and PRISM will be able to probe the CMB distortions predicted by $g_{AX}={\cal O}(1)$ which is most favored in our scenario. We note also that for the fuzzy dark matter with $m_\phi\sim 10^{-21}$ eV \cite{Hu:2000ke}, the resulting value of $\Delta N_{\rm eff}$ is close to the bound from CMB observation \cite{Ade:2015xua}, which might be an interesting point in connection with the discrepancy in the values of $H_0$  inferred from  CMB data and local measurements.

Dark photon fields may contribute to metric perturbations as well. While their characteristic scale 
$k_*$ is beyond the reach of direct cosmological probes 
(e.g. Lyman-$\alpha$ forests), they can source the acoustic oscillation of photon baryon fluid and contribute to the CMB spectral distortion\,\cite{Chluba:2011hw,Khatri:2012tw}, which deserves more detailed study \cite{detail}.
 
Our scenario assumes $g_{XX}={\cal O}(10-100)$ and $g_{AX} = {\cal O}(1)$. One may ask whether such ALP effective couplings can be obtained from a sensible UV completion of the model. 
If one assumes that the field range of periodic ALP is of ${\cal O}(2\pi f)$, then naive field theoretic consideration suggests  that 
$g_{XX}={\cal O}(e_X^2/8\pi^2)$ and $g_{AX}={\cal O}(ee_X/8\pi^2)$, which appear to be significantly smaller than 
the values assumed in our scenario.
 This problem can be easily solved  by the clockwork mechanism \cite{Choi:2014rja,Choi:2015fiu,Kaplan:2015fuy},  enlarging the ALP field range exponentially, while keeping the ALP couplings to gauge fields essentially fixed.  An explicit realization along this direction will be presented in the forthcoming  paper\,\cite{detail}, together with more extensive study of our magnetogenesis scenario.

\begin{acknowledgments}
This work was supported by IBS under the project code, IBS-R018-D1.
T.S. is supported by by JSPS KAKENHI Grant Numbers JP15H02082, 18H04339, 18K03640.
K.C. thanks K. Kamada, S. Lee and H. Seong, and T.S. thanks J. Yokoyama, M. Oguri and T. Shigeyama for fruitful discussions.  
\end{acknowledgments}

\bibliography{magnetogenesis}
\end{document}